\documentclass[manuscript]{aastex631_arxiv}

\usepackage{color}
\usepackage{graphicx}
\usepackage{bm}
\usepackage{epsf}
\usepackage{wrapfig}
\usepackage{multirow}
\usepackage{epstopdf}
\usepackage{tabularx}
\usepackage{amssymb,amsmath}

\shorttitle{solar cycle: different maps}
\shortauthors{Huang et al.}

\begin{document}

\title{Solar Wind Driven from GONG Magnetograms in the Last Solar Cycle}

\correspondingauthor{Zhenguang Huang}
\email{zghuang@umich.edu}

\author{Zhenguang Huang}
\affiliation{Climate and Space Sciences and Engineering, University of Michigan, Ann Arbor, MI 48109, USA}

\author{G\'abor T\'oth}
\affiliation{Climate and Space Sciences and Engineering, University of Michigan, Ann Arbor, MI 48109, USA}

\author{Nishtha Sachdeva}
\affiliation{Climate and Space Sciences and Engineering, University of Michigan, Ann Arbor, MI 48109, USA}

\author{Bart van der Holst}
\affiliation{Climate and Space Sciences and Engineering, University of Michigan, Ann Arbor, MI 48109, USA}

\begin{abstract}
In a previous study, \cite{Huang_2023} used 
the Alfv\'en Wave Solar atmosphere Model (AWSoM), one of the widely
used solar wind models in the community, driven by 
ADAPT-GONG magnetograms to simulate the solar wind in the last solar cycle
and found that the optimal Poynting flux parameter 
can be estimated from 
either the open field area or
the average unsigned radial component of the magnetic field in the open field regions.
It was also found that the average energy deposition rate (Poynting flux) 
in the open field regions is approximately constant. In the current study,
we expand the previous work by using GONG magnetograms
to simulate the solar wind for 
the same Carrington rotations
and determine if the results are similar to the ones obtained with ADAPT-GONG magnetograms. 
Our results indicate that 
similar correlations can be obtained from the GONG maps. 
Moreover, we report that
ADAPT-GONG magnetograms can consistently provide better comparisons
with 1\,AU solar wind observations than GONG magnetograms, based on the
best simulations selected by the minimum of the average curve distance for the solar wind speed and density.

\end{abstract}

\section{Introduction}

Accurately predicting the solar wind distribution at various locations in the heliosphere is one of the major objectives 
in the solar and heliosphere community, as well as the space weather community. 
The background solar wind plays a crucial role in determining the strength and arrival time of various
drivers of space weather events, e.g., Coronal Mass Ejections (CMEs), Solar Energetic Particles (SEPs) 
and co-rotating interaction regions (CIRs). CIRs or CMEs can cause large
geomagnetic disturbances (geomagnetic storms) that threaten advanced technology that we are
highly reliant on. For example, the 1989 geomagnetic storm caused a widespread effect on the power grid system
including a blackout of the Hydro-Qu\'ebec system \citep{Boteler_2019}.

There are many different solar wind models in the community, from empirical or semi-empirical models like the 
commonly used Wang-Sheeley-Arge (WSA) model \citep{Wang_1990, Wang_1992, Arge_2000}, 
to first principles models focusing on small scale reconnections \citep{Axford_1992, Fisk_1999, Fisk_2001, Schwadron_2003, Fisk_2003, Yamauchi_2004} as well as plasma wave heating \citep{Tu_1997, Mikic_1999, Hu_2000, Usmanov_2000, Dmitruk_2002, Markovskii_2002, Li_2003, Vasquez_2003, Suzuki_2005, Cohen_2007, Cranmer_2010, Feng_2011, Verdini_2010, Osman_2011, Chandran_2011, Lionello_2014b}, 
and the newly developed machine learning model by \cite{Upendran_2020}. All these models make different assumptions, 
and/or apply different methods to predict the solar wind. But there is (at least) one thing in common: all of them need
observational input(s) from the Sun, either the observed magnetic field on the surface, or the 
solar corona ultraviolet image(s).

The photospheric magnetic field is the primarily used observational input for solar wind models.
There are two major types of observational products: synoptic magnetograms, which reconstruct
the map from 27-day of magnetic field observations with some simple weighting factors 
(e.g., the Global Oscillation Network Group \cite[GONG,][]{Harvey_GONG_1996}
uses $\cos^4$ of longitude) 
to ensure measurements taken at a particular  time contribute 
most to that Carrington longitude in the synoptic map; and  
synchronic magnetograms, which take the photospheric magnetic field observations and 
apply surface flux transport models to
simulate the evolution of the surface magnetic field 
while assimilating new observations. 
There are different observational sources used in both types of magnetograms, e.g., the Helioseismic
Magnetic Imager \cite[HMI,][]{Schou_HMI_2012} on the Solar Dynamics Observatory (SDO), 
the Michelson Doppler Imager \cite[MDI,][]{Scherrer_MDI_1995} on the Solar and Heliospheric Observatory (SoHO), 
the GONG, the Mt. Wilson Observatory \cite[MWO,][]{Ulrich_MWO_2002}, 
the Synoptic Optical Long-term Investigations of the Sun 
\cite[SOLIS,][]{Keil_SOLIS_2003} at the National Solar Observatory (NSO), and 
the Wilcox Solar Observatory \cite[WSO,][]{Babcock_1953,Scherrer_1977}. Also, there are different flux transport models, e.g., 
the Air Force Data Assimilative Photospheric flux Transport (ADAPT, \cite{Hickmann_2015}), 
the Advective Flux Transport (AFT, \cite{Upton_2014a, Upton_2014b}), and
the Lockheed Martin flux transport model \citep{Schrijver_2003}.

The input magnetogram has a strong impact on the simulated solar wind.
\cite{Gressl_2014} used four different 
magnetograms (MDI, NSO, MWO and GONG) to simulate the solar wind and 
showed that the choice of the synoptic map for the solar wind models
significantly affects the model performance. \cite{Jian_2015} compared multiple coronal
and heliospheric models and different magnetograms and arrived at similar conclusions. 
Recently, \cite{Sachdeva_2019} used the Alfv\'en Wave Solar atmosphere Model (AWSoM)
to simulate solar minimum conditions and they noticed that the simulated solar wind
based on the ADAPT-GONG magnetogram agrees better with observations, than the GONG magnetogram when all other model parameters are kept the same. 
\cite{Jin_2022} assessed the influence of input magnetic maps on modeling the solar wind
and concluded that it is important to consider the model uncertainty due to the imperfect magnetic field measurements.
\cite{Sachdeva_2021,Sachdeva_2023} used AWSoM to simulate solar maximum conditions with different input
magnetograms, and confirmed that the simulated solar wind 
at 1\,AU can be very different when different input magnetograms are used.

In this study, we will expand our previous work (\cite{Huang_2023}, Paper I, hereafter) by simulating the solar
wind using GONG magnetograms, for the same Carrington rotations that Paper I
studied using ADAPT-GONG magnetograms. We then investigate how 
GONG maps affect the simulated solar wind at 1\,AU
during different phases of the last solar cycle. 
Both ADAPT-GONG and GONG
magnetograms are widely used as the input magnetic field map for solar wind 
modeling. They take the same observations but construct the maps using different
techniques. ADAPT-GONG applies the ADAPT surface 
flux transport model while GONG takes simple weighting factors. 
Systematically evaluating the model 
performance when driven by the two different maps
can provide insights if a particular type of map provides
better simulated solar wind, which is critical
for real-time solar wind prediction. 
We will use AWSoM \citep{Sokolov_2013,vanderholst_2014,Sokolov_2021}
to simulate the steady-state background solar wind. 
AWSoM is one of the widely used solar wind models in the community. It is
open-source on GitHub (\url{https://github.com/MSTEM-QUDA/SWMF}) and 
can also be accessed via the runs-on-request service provided by the 
Community Coordinated Modeling Center (CCMC). AWSoM has been 
extensively validated with in-situ and remote observations under
different solar wind conditions \citep{Jin_2012, Oran_2013, Sachdeva_2021, Huang_2023, Szente_2022, Szente_2023}, 
as well as at different heliocentric distances including Parker Solar Probe
locations \citep{vanderholst_2019, vanderholst_2022}.

\section{Methodology} 

AWSoM is implemented in the BATS-R-US (Block Adaptive Tree Solar Wind Roe-type
Upwind Scheme) code \citep{Groth_2000, Powell_1999} within the Space Weather Modeling Framework (SWMF) \citep{Toth_2005, Toth_2012, Gombosi_2021}. 
The main input of the model is the observed radial component of the photospheric 
magnetic field, a magnetogram, at the inner boundary. 
The density and temperature at the inner boundary are uniformly specified  as 
$n=2\times10^{17}$\,m$^{-3}$
and $T=50,000$\,K. 
AWSoM assumes that the pressure gradient and the nonlinear dissipation of the Alfv\'en wave turbulence
are the only sources for accelerating and heating the solar wind, respectively. 
At the outer boundary, AWSoM applies a zero gradient condition 
allowing the super-fast magnetosonic solar wind to freely leave the domain. 
In depth discussions of the physics can be found in 
\cite{Sokolov_2013,vanderholst_2014,vanderholst_2022}.

Significant progress has been made in understanding how the input parameters of AWSoM
affect the simulation results recently. \cite{Jivani_2023} used uncertainty quantification to study the effect of using different model parameters and determined
the three major parameters impacting the simulated output. These are the Poynting flux parameter, associated with the energy input 
at the inner boundary for heating the corona and driving the solar wind; 
the perpendicular correlation length parameter, associated with how solar wind plasma gains energy from 
the Alfv\'en wave turbulence in the simulation domain; and the multiplicative factor 
for the input magnetogram, which is related to the uncertainty of the magnetic field observations.
Paper I focused on the variation of the optimal value for the Poynting flux parameter in the last solar cycle
and determined that it is 
linearly correlated with the area of the open field regions and anti-correlated with
the average unsigned radial component of the magnetic field.

In this study, we apply the same approach as Paper I. We simulate the same 
Carrington rotations as Paper I, which are listed in Table~\ref{tab:cr}, 
so that we can have a direct comparison between simulation results using the ADAPT-GONG and GONG maps. 
The ADAPT-GONG magnetograms are available at \url{https://gong.nso.edu/adapt/maps/gong},
and the GONG magnetograms can be accessed on \url{https://gong.nso.edu/data/magmap/}. 
We apply a similar approach as in Paper I and vary the the Poynting flux parameter  between 0.3 and 1.2$\,\mathrm{MWm}^{-2}\mathrm{T}^{-1}$ with every 0.05$\,\mathrm{MWm}^{-2}\mathrm{T}^{-1}$, except for two rotations, CR2137 and CR2198, 
in which the optimal values are outside
the typical range. For CR2137, the range is extended below 0.3$\,\mathrm{MWm}^{-2}\mathrm{T}^{-1}$ by adding extra
points as [0.1, 0.125, 0.15, 0.175, 0.2, 0.25]$\,\mathrm{MWm}^{-2}\mathrm{T}^{-1}$; and for CR2198, three extra
points of [1.25, 1.3, 1.35]$\,\mathrm{MWm}^{-2}\mathrm{T}^{-1}$ 
are included.
All other parameters are set with the default values. To be 
specific, the perpendicular correlation length parameter is set to 
1.5$\times 10^5\,\mathrm{m\,T}^\frac{1}{2}$. 
For the GONG magnetograms, we apply a scaling factor of 3.75 to account for
the uncertainty of the synoptic magnetogram observations,
especially the very weak field in high latitude regions.
To be specific, we set the radial component of the magnetic field ($B_r$) to sign($B_{r\text{GONG}}$)*min($3.75 \cdot \lvert B_{r\text{GONG}} \rvert $, 
$5 + \lvert B_{r\text{GONG}} \rvert$), 
where $B_{r\text{GONG}}$ is the observed $B_r$ from the GONG magnetogram and the unit is in Gauss. 
This setting is consistent with previous studies \citep{Cohen_2007,Jin_2012}. 
\cite{Riley_2007} also applies a similar factor to study the open flux problem.
As \cite{Hickmann_2015} applied their scaling factor 
to derive the ADAPT-GONG magnetogram,
we do not apply such a multiplicative factor for ADAPT-GONG magnetograms. 
There are 180 simulations in total in this study. Each simulation takes about 4000 to 5000 CPU hours (1120 cores running for about 3.5-4.5 hours) for the Intel 8280 ``Cascade Lake" type CPU, leading to
a total cost of about 800K CPU hours.

We use the curve distance, introduced by \cite{Sachdeva_2019}, between the simulation output and in situ observations at 1\,AU to
quantify the model performance. We select the optimal value of the Poynting flux parameter
when the average of the curve distances of the solar wind density and velocity reaches
its minimum, the same as Paper I.

\begin{table}[]
\center
\begin{tabular}{|c|cc|}
\hline
\multirow{2}{*}{Carrington Rotation} & \multicolumn{2}{c|}{UTC Time of the Maps}                                     \\ \cline{2-3} 
                                     & \multicolumn{1}{c|}{ADAPT-GONG} & GONG    \\ \hline
2106                                 & \multicolumn{1}{c|}{2011-2-2  02:00}      & 2011-2-1  23:54  \\ \hline
2123                                 & \multicolumn{1}{c|}{2012-5-16 20:00}     & 2012-5-16 17:54  \\ \hline
2137                                 & \multicolumn{1}{c|}{2013-5-28 20:00}     & 2013-5-28 20:04  \\ \hline
2154                                 & \multicolumn{1}{c|}{2014-9-2 20:00}       & 2014-9-2 20:04   \\ \hline
2167                                 & \multicolumn{1}{c|}{2015-8-23 02:00}     & 2015-8-23 02:04  \\ \hline
2174                                 & \multicolumn{1}{c|}{2016-3-3 02:00}       & 2016-3-3 02:14   \\ \hline
2198                                 & \multicolumn{1}{c|}{2017-12-17 02:00}    & 2017-12-17 02:04 \\ \hline
2209                                 & \multicolumn{1}{c|}{2018-10-13 06:00}    & 2018-10-13 06:04 \\ \hline
2222                                 & \multicolumn{1}{c|}{2019-10-2 02:00}      & 2019-10-2 02:14  \\ \hline
\end{tabular}
\caption{All the ADAPT-GONG and GONG magnetograms used in this study. For all the ADAPT-GONG maps, the
7th realization is picked as mentioned in Paper I.}
\label{tab:cr}
\end{table}

\section{Simulation Results} 

\begin{figure}[ht!]
\center
\gridline{\fig{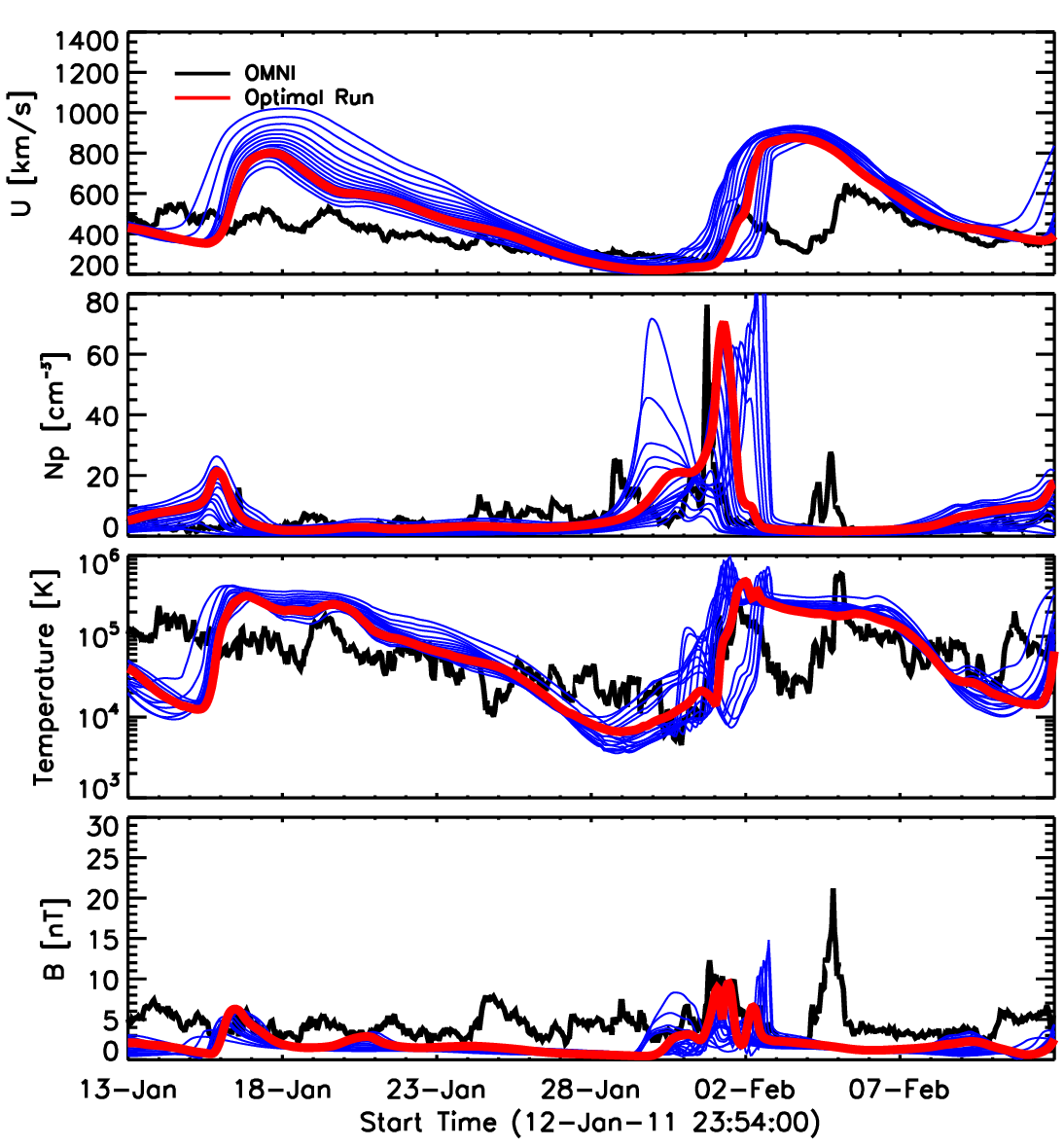}{0.48\textwidth}{(a) GONG}
          \fig{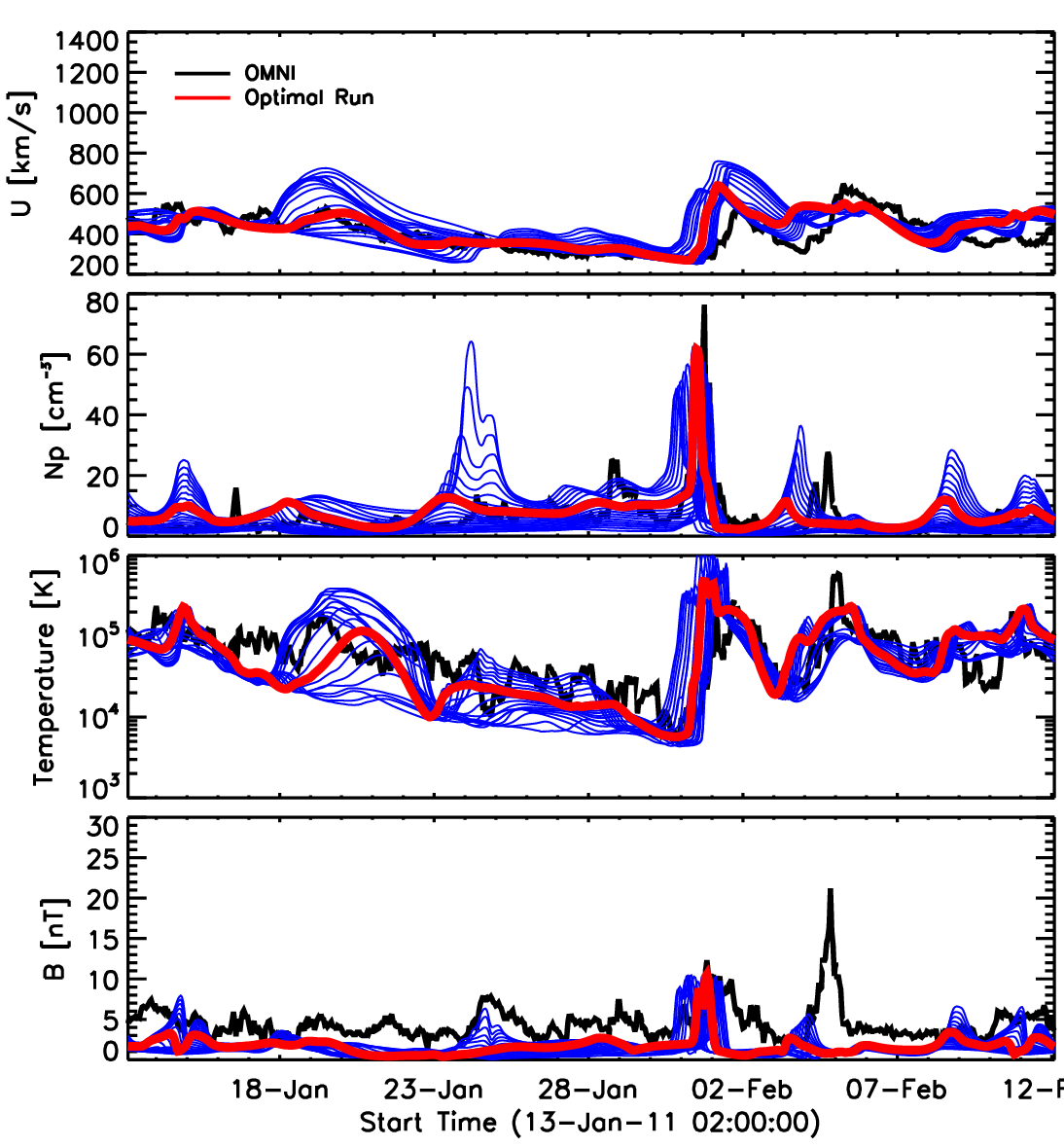}{0.48\textwidth}{(b) ADAPT-GONG}}
          
\caption{Comparison between the AWSoM simulated solar wind (blue and red lines) and the hourly OMNI data (black lines) at 1\,AU, for CR2106, which
is near solar minimum in 2011. Each blue line is associated with a specific value of the Poynting flux parameter, 
and the red line highlights the simulated solar wind using the optimal Poynting flux parameter.
Panel (a) is based on the GONG magnetogram, while Panel (b) is a reproduction based on the ADAPT-GONG magnetogram for the same Carrington Rotation from Paper~I. 
}
\label{fig:diff_pf_2106}
\end{figure}

\begin{figure}[ht!]
\center
\gridline{\fig{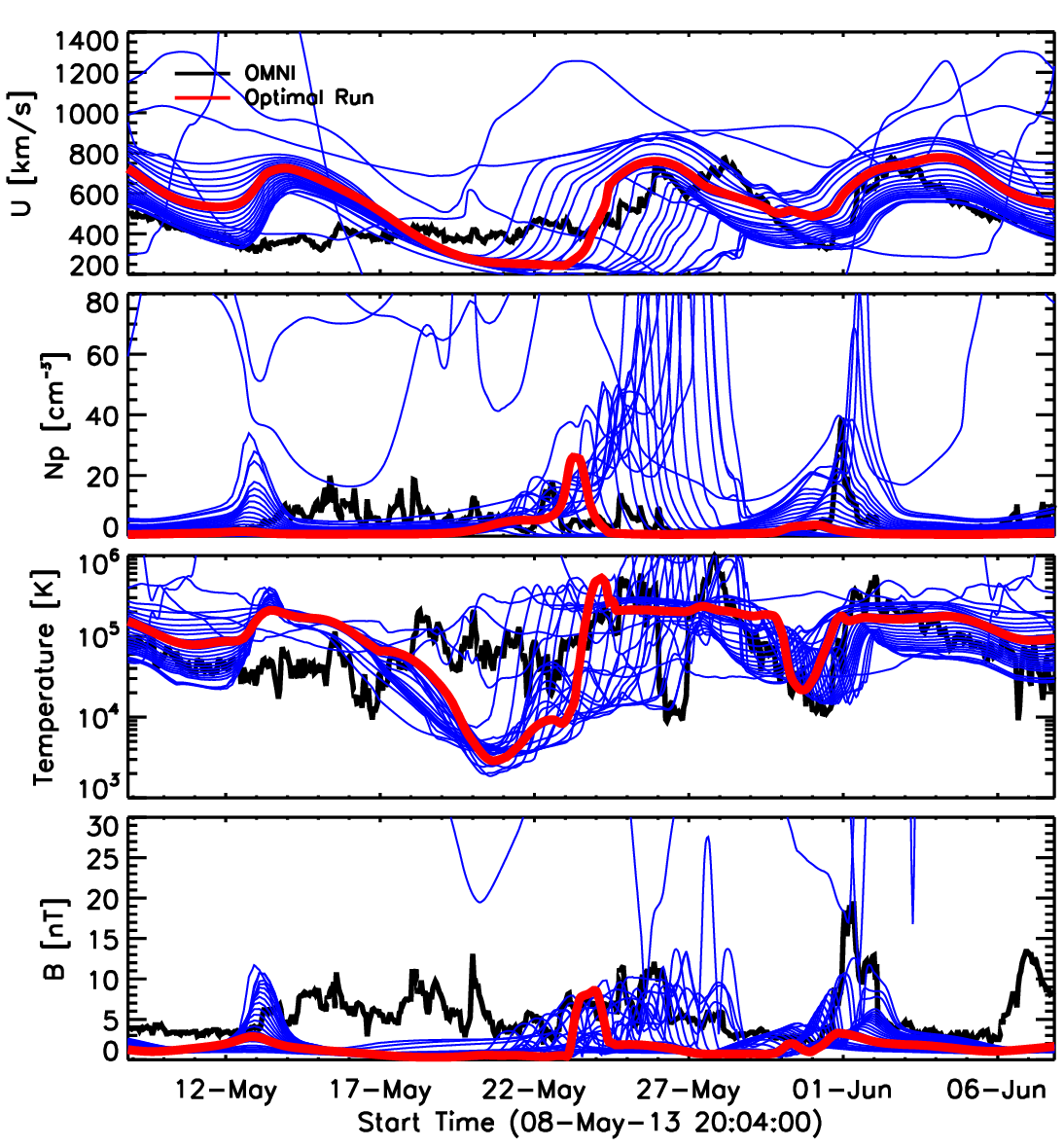}{0.48\textwidth}{(a) GONG}
          \fig{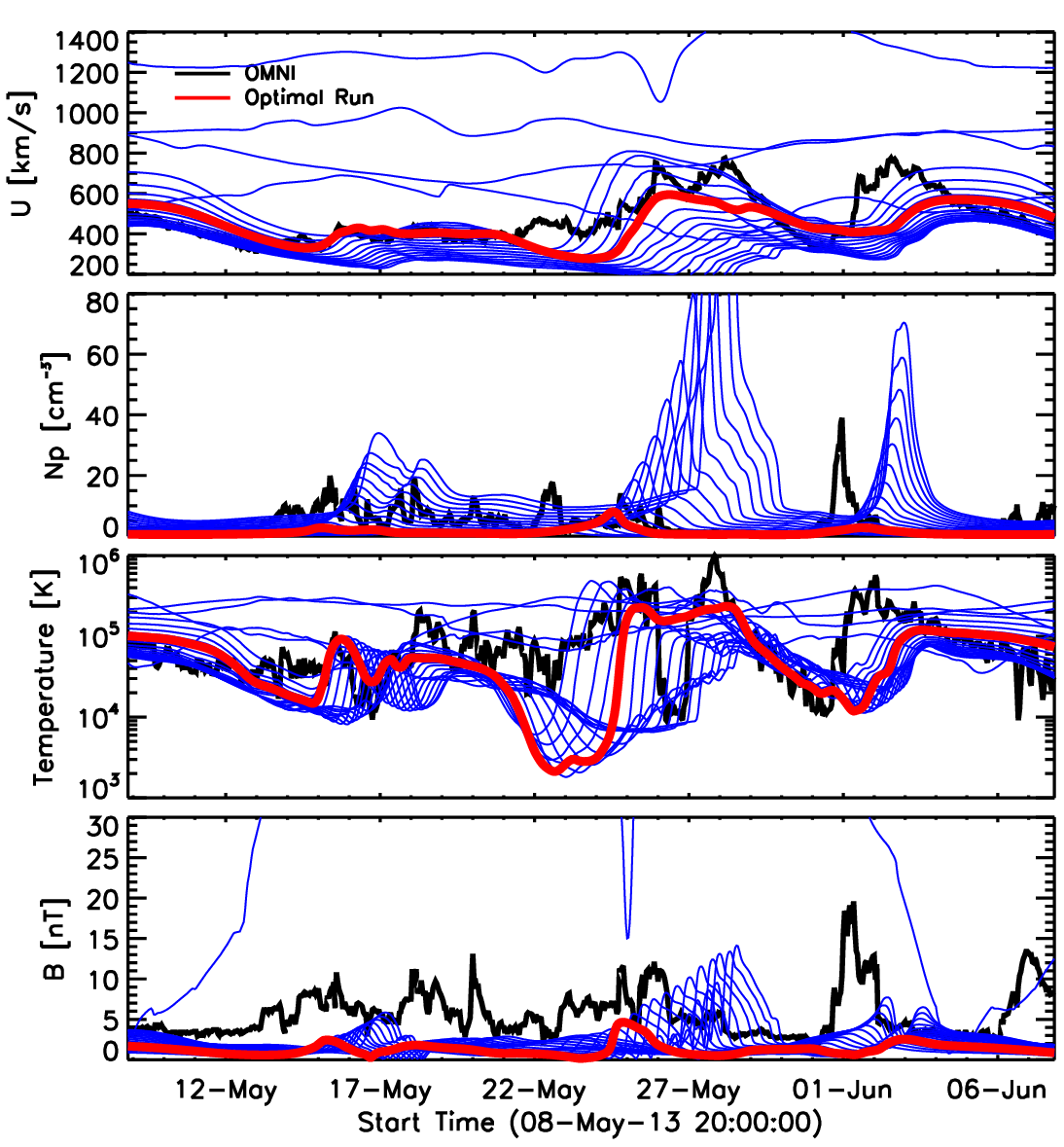}{0.48\textwidth}{(b) ADAPT-GONG}}
          
\caption{This figure shows a similar comparison as Figure~\ref{fig:diff_pf_2106} but for CR2137, which is near solar maximum in 2013. Panel (a) is based on the 
GONG magnetogram, while Panel (b) is a reproduction based on the ADAPT-GONG 
magnetogram from Paper I.
}
\label{fig:diff_pf_2137}
\end{figure}

For clear one-to-one comparisons between the current study and our previous results using ADAPT-GONG maps (Paper I), we show figures similar to Paper I.
We plot the AWSoM simulated solar wind (driven by GONG synoptic maps) for the same Carrington rotations:
CR2106 near solar minimum in Figure~\ref{fig:diff_pf_2106}, and CR2137 near solar maximum in Figure~\ref{fig:diff_pf_2137}. Because
we only change the values of the Poynting flux parameter of AWSoM, this figure clearly demonstrates the variations in the simulated solar wind results at 1\,AU due to the different Poynting flux parameters. 
Similar to Figure~1 in Paper I, the simulations with different Poynting flux parameters are shown in blue and
the optimal value of the Poynting flux parameter is highlighted in red. These figures share some similarities:
the variations of the solar wind due to different Poynting flux parameters are smaller near solar minimum (CR2106)
and larger near solar maximum (CR2137); and some Poynting flux parameters produce unphysical 
results for a relatively active Sun (CR2137). But there are some distinct 
differences: 
the simulated solar wind bulk velocity 
based on the GONG magnetograms (Panels (a) in Figures~\ref{fig:diff_pf_2106} and
\ref{fig:diff_pf_2137}) is significantly larger than
the ADAPT-GONG magnetograms
(Panels (b) in Figures~\ref{fig:diff_pf_2106} and
\ref{fig:diff_pf_2137}), while the density is about the same or slightly smaller. 

\begin{figure}[ht!]
\center
\gridline{\fig{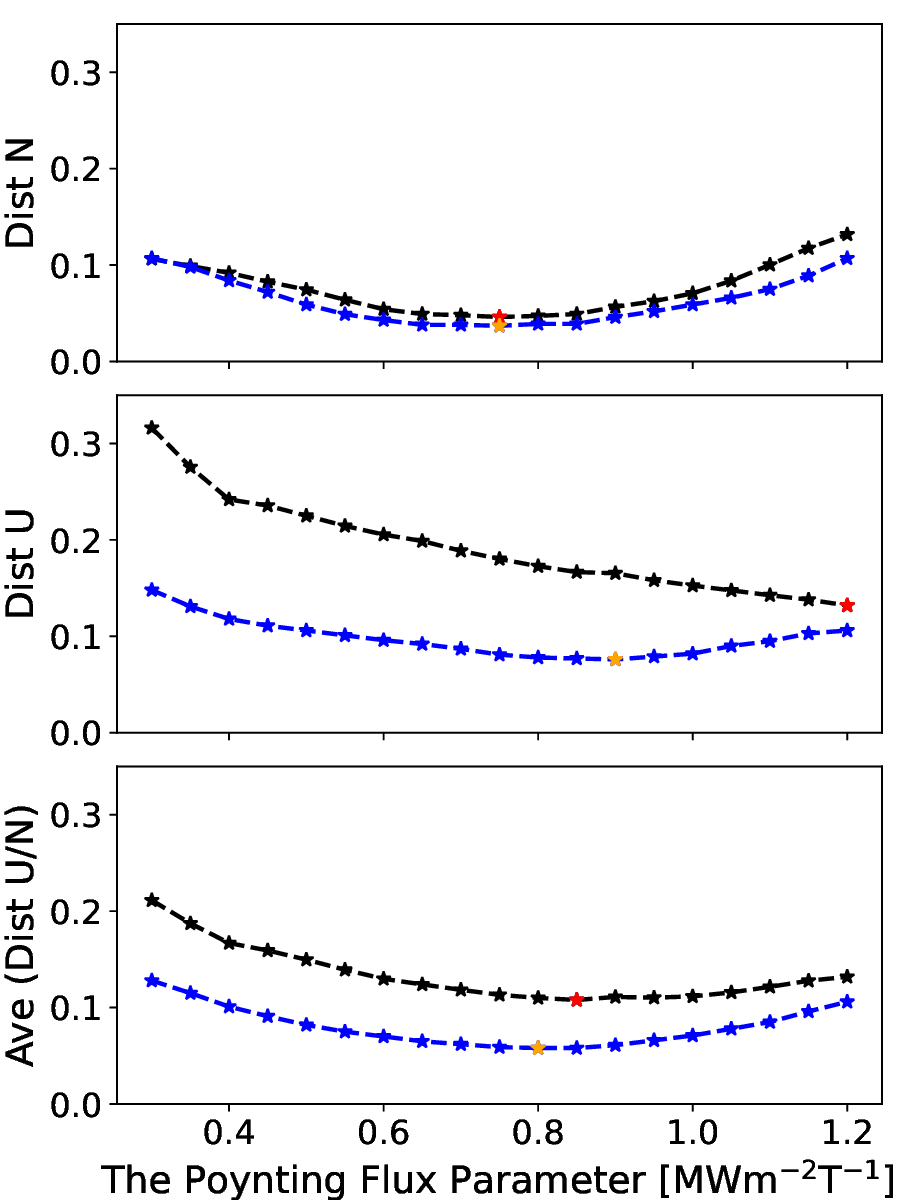}{0.45\textwidth}{(a)}
          \fig{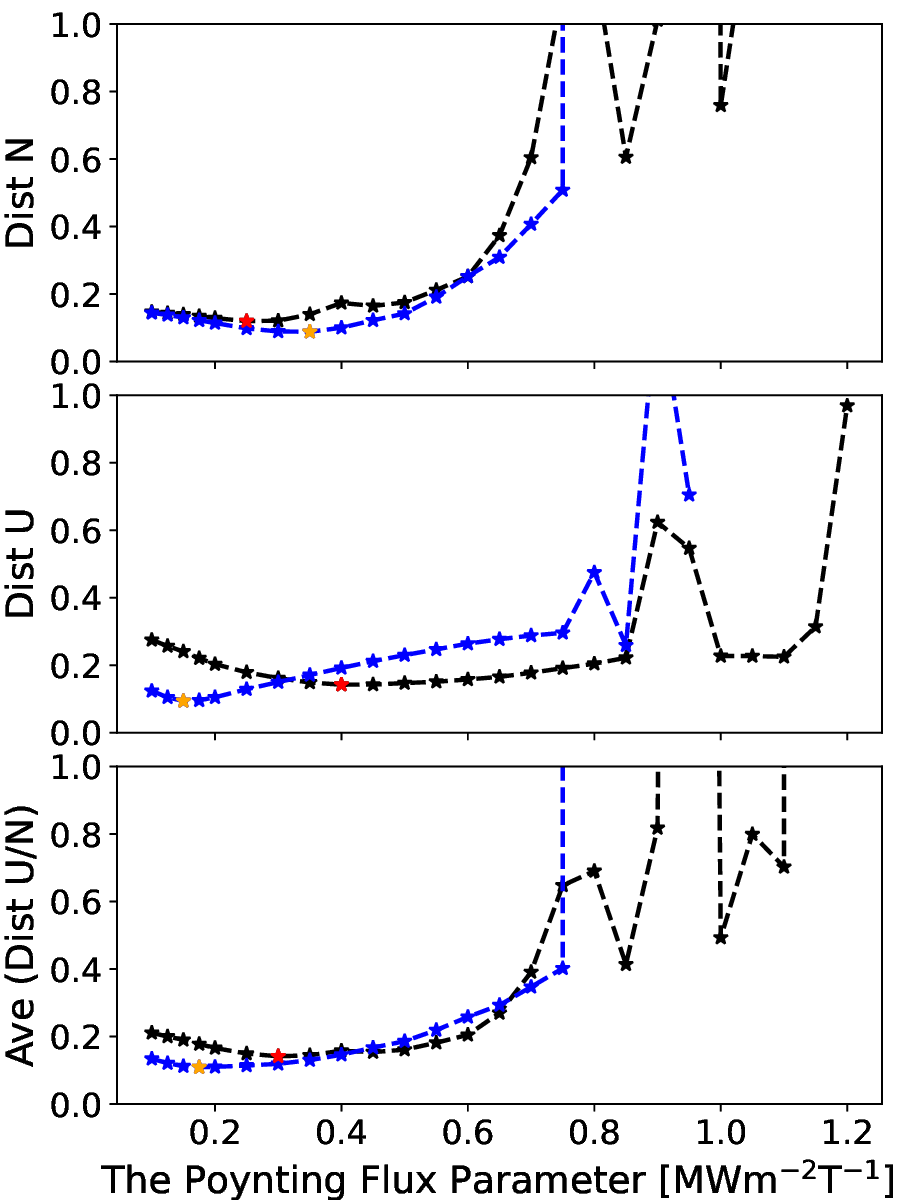}{0.45\textwidth}{(b)}}
          
\caption{The variations of the distances as a function of different
Poynting flux parameters. CR2106 is shown in Panel (a) while
CR2137 is plotted in Panel (b). The relative curve distances
for the solar wind density and velocity, as well as the average value
of velocity and density distances are displayed from top to bottom, 
respectively. The black color is associated with the GONG magnetograms
with the red star highlighting the optimal Poynting flux parameter value for
the corresponding sub-figure, 
while the blue color is reproduced from Paper~I for the
ADAPT-GONG magnetogram results, with the yellow star highlighting the optimal Poynting flux parameter value.
}
\label{fig:dist}
\end{figure}

Paper I used the curve distance formula introduced by \cite{Sachdeva_2019} to quantify the solar
wind variations due to different Poynting flux parameters. We use the identical method in this study
for a direct comparison. Figure~\ref{fig:dist} show the same quantities as Figure~2 in Paper I:
the curve distance between the simulated solar wind density (top), bulk velocity (middle), and the
average of the density and velocity distances (bottom) for the GONG magnetograms (in black) and ADAPT-GONG magnetograms (in blue), with the
red and orange stars highlighting the local optimal Poynting flux parameter for the corresponding sub-figure.
The optimal Poynting flux parameter is selected when the average of the density and velocity distances
reaches its minimum, which is the same criteria as applied in Paper I.
Similar features can be identified between GONG and ADAPT-GONG magnetograms.  
The simulated solar wind becomes unphysical when the Poynting flux parameter
is larger than a certain value for a CR2137 (near solar maximum). The threshold is around 
0.7\,MWm$^2$T$^{-1}$ in this study, slightly smaller than the 0.75 \,MWm$^2$T$^{-1}$ 
for the ADAPT-GONG magnetogram in Paper~I. 
We also identify a monotonic increase of the distance when the Poynting flux 
parameter is smaller or larger than the local optimal value, 
except for the distance value plot for the solar wind bulk 
velocity for CR2106 in Panel~(a).
The curve distance for the solar wind bulk velocity for CR2106 is monotonically
declining. It is unclear if it has reached its minimum yet. 
But because we select the optimal value based on the average
of the velocity and density curve distances, 
the optimal value is already identified. To save computational cost,
we do not extend the range of Poynting flux parameter
for this rotation to confirm if the distance for the velocity 
will increase after it reaches its minimum.
The monotonic trend is also identified in Paper I, which again suggests that 
the selection of the optimal value is reliable.
 Moreover, compared to the ADAPT-GONG results reproduced from Paper~I, we notice 
larger curve distances when GONG magnetograms are used, 
which is consistent with the discussion above. 

In order to better quantify the differences between the curve distances obtained from ADAPT-GONG and GONG magnetograms,
we list the minimum values of the average distance of the solar wind density and bulk velocity in Table\,\ref{tab:dist}. We confirm
that for most of the rotations in the study (8 out of 9), the optimal runs based on ADAPT-GONG magnetograms give
smaller minimum distance values than the GONG magnetograms, with only one exception for CR2154 when the GONG magnetogram provides a slightly better result than the ADAPT-GONG magnetogram. 

\begin{figure}[ht!]
\center
\gridline{\fig{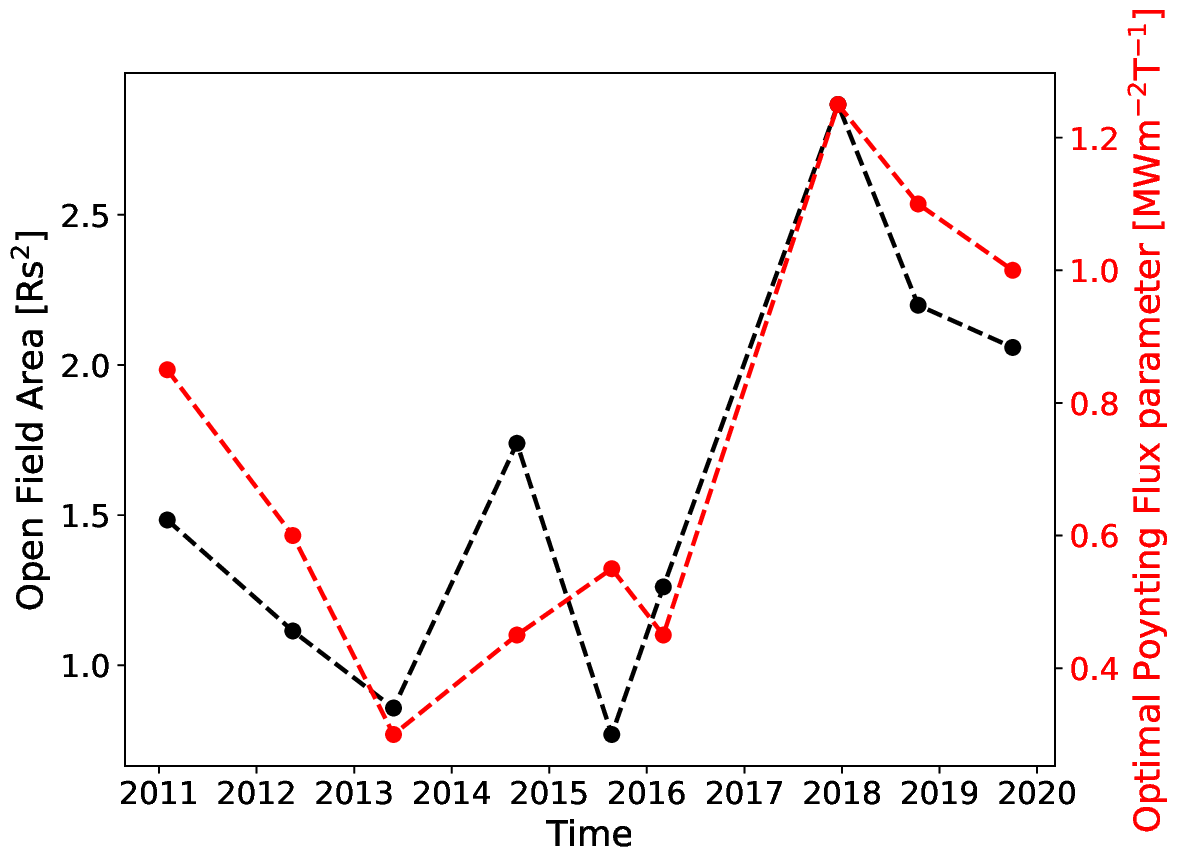}{0.49\textwidth}{(a)}
          \fig{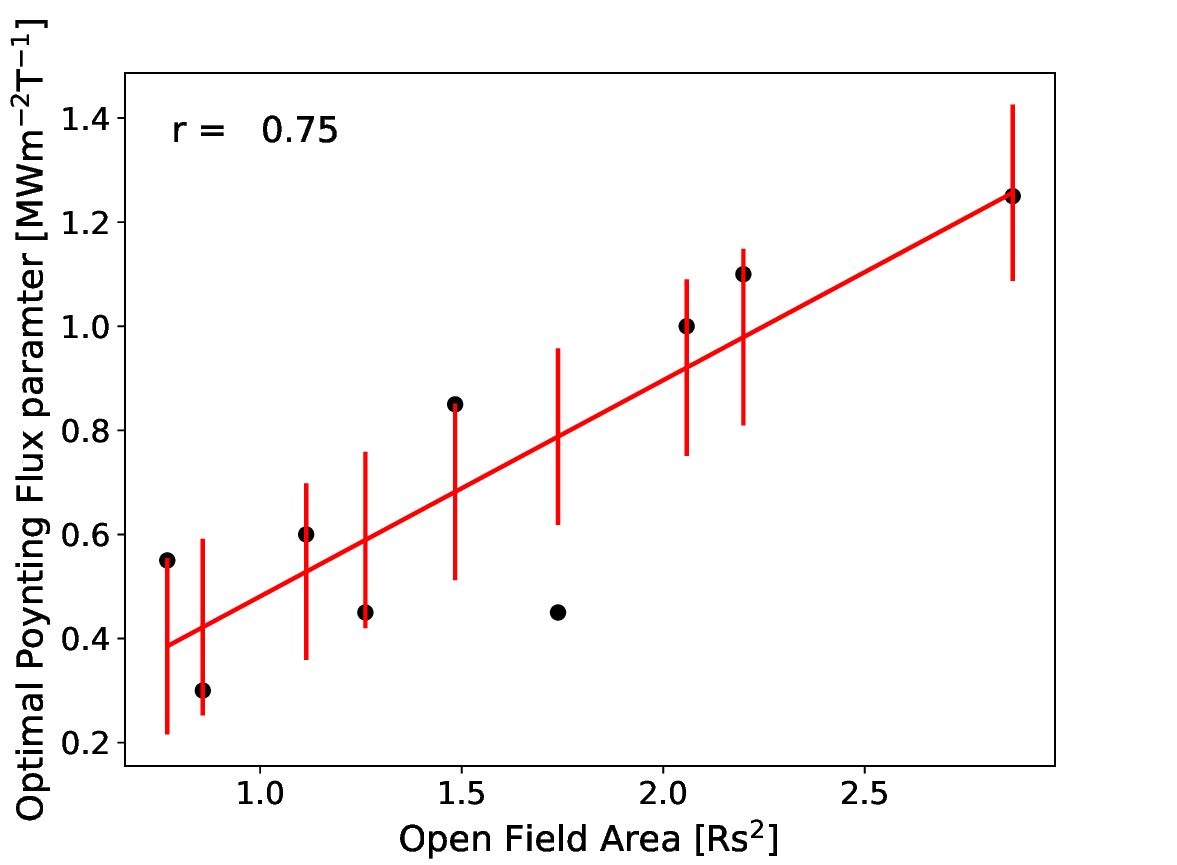}{0.49\textwidth}{(b)}}
\gridline{\fig{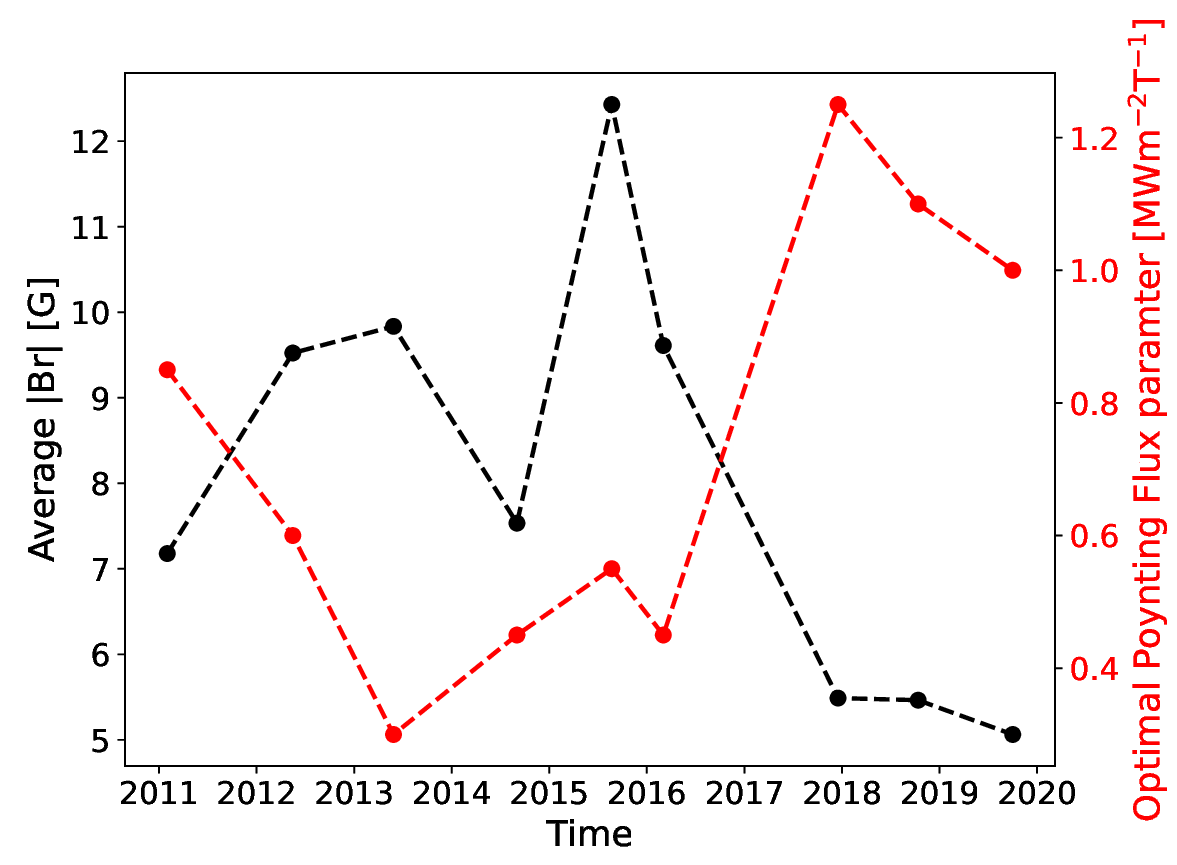}{0.49\textwidth}{(c)}
          \fig{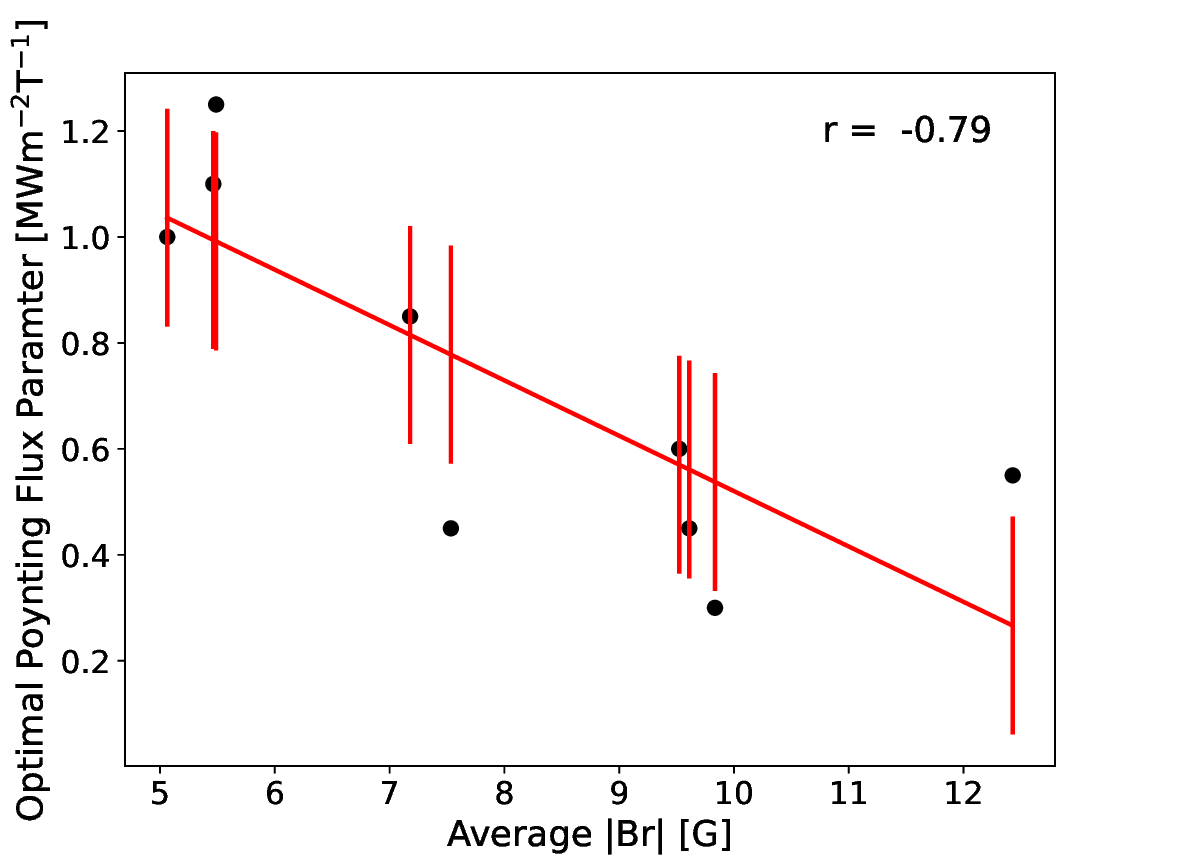}{0.49\textwidth}{(d)}}
\caption{The correlations between the optimal Poynting flux parameter
and the open field area as well as the average unsigned $B_r$ in the
open field regions. Panels\,(a) and (c) show the time evolution of
the optimal Poynting flux parameter (in red), 
the open field area (in black) and the average unsigned $B_r$ (in black). 
Panels\,(b) and (d) plot the linear
regression results between the optimal Poynting flux parameter and
the open field area and the average unsigned $B_r$, respectively, 
with the Spearman's correlation shown in the upper left corner. 
The error bars indicate the standard deviation of the fit.}
\label{fig:stat}
\end{figure}

Next, following Paper I, we examine if the optimal Poynting flux parameter has any relation with the area of the open field regions and/or the average unsigned radial component of the magnetic field ($|B_r|$). The results are presented in Figure~\ref{fig:stat}. We conclude that similar relations exist for simulations using GONG magnetograms as the one using ADAPT-GONG. However, the Spearman's correlation coefficients $r$ for the GONG magnetograms are smaller than for the ADAPT-GONG magnetograms: the coefficient for the area of open field regions is 0.75, compared to 0.96 obtained with ADAPT-GONG magnetograms; and the coefficient for average unsigned $|B_r|$ in the open field regions is $-0.79$, compared to $-0.91$ found with ADAPT-GONG magnetograms. Similarly, we obtain the following formulas from the linear regression:
\begin{eqnarray}
P &=& 0.42  \cdot A + 0.07 \pm 0.17 \\
P &=& -0.1 \cdot B + 1.57 \pm 0.21
\end{eqnarray}
where $P$ is optimal Poynting flux parameter in the unit of [MWm$^{-2}$T$^{-1}$], $A$ is the area of the open field regions in the unit of [R$_s^2$], and $B$ is the average unsigned  $B_r$ in the open field regions in the units of [G]. The standard error of the linear regression is given by the $\pm$ terms. Compared to Equations\,1 and 2
in Paper I, the slope for the open field area is the same while it is
slightly a little larger for the average unsigned $B_r$ in the open regions (-0.07 in Paper I). The intersections and standard errors
for both equations are larger than the values in Paper I: the intersection/standard errors is 0.02/0.11 for the open field area and 1.29/0.16 for the average unsigned $B_r$ in Paper I. The larger standard
errors for GONG magnetograms result from larger variations when
using different Poynting flux parameters as compared to ADAPT-GONG based simulations.

\begin{figure}[ht!]
\center
\includegraphics[width=0.9\linewidth]{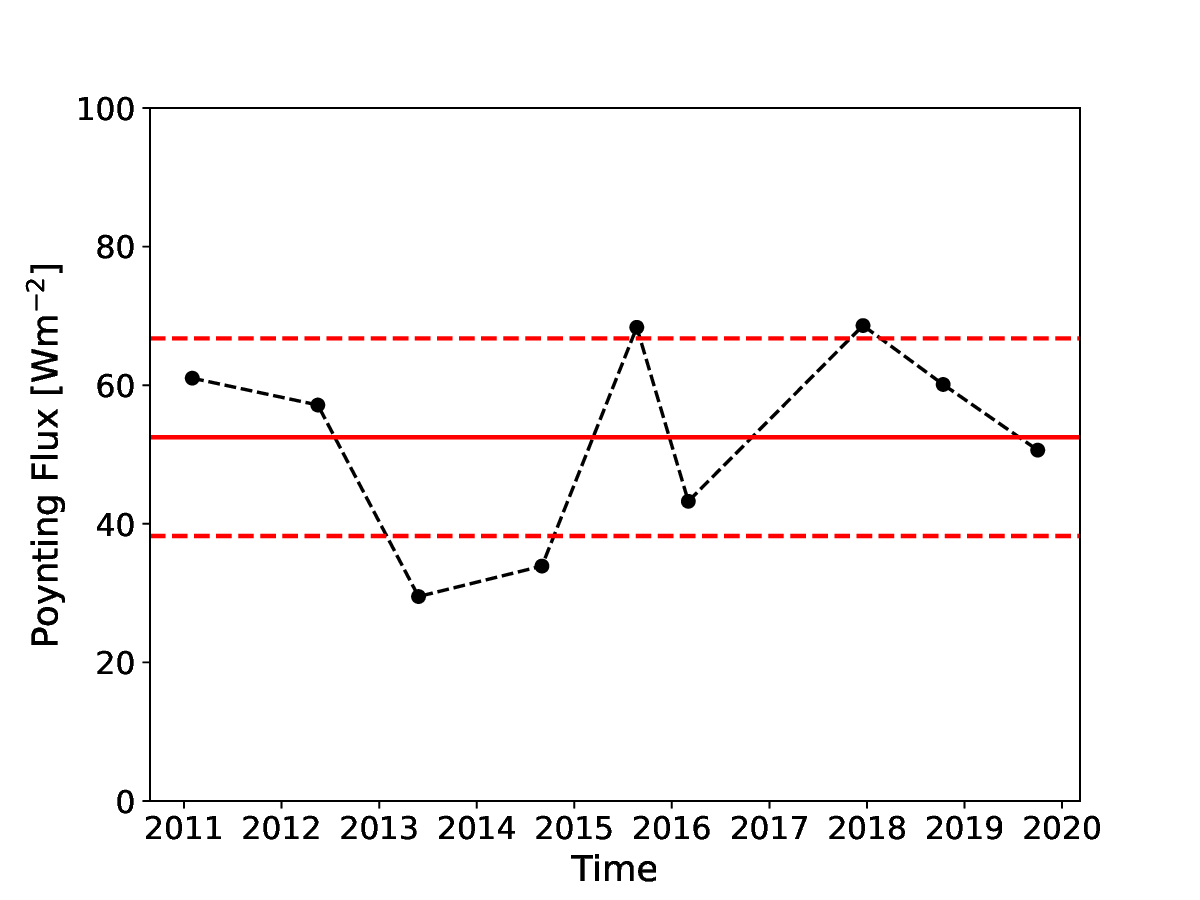}
\caption{The time evolution of the average energy deposition rate
(Poynting flux) in the open field regions between 2011 and 2019. The
average value (52.5\,Wm$^{-2}$) is shown with the horizontal red
line while the standard deviation (14.3\,Wm$^{-2}$) is
plotted with the horizontal red dashed lines.
}
\label{fig:PoyntingFlux}
\end{figure}

Paper I discovered that the average energy deposition rate (the Poynting flux) in 
the open field regions does not show significant variations among the simulated CRs. In order
to further understand if this is true for the GONG magnetograms, 
we calculate the same quantity, which is shown 
in Figure~\ref{fig:PoyntingFlux}. This figure is very similar to Figure~4 
in Paper I. The average Poynting flux in the open field regions 
is found to be $52.5\pm14.26\,$Wm$^{-2}$, while it is $47.42\pm13.12\,$Wm$^{-2}$ 
for the ADAPT-GONG magnetograms in Paper I. 
They are essentially the same within the range of the standard error. 
To conclude, our study confirms that this behavior (small variations in the average deposition rate over time) is also valid when using the GONG magnetograms as the input for AWSoM.

\begin{table}[]
\center
\begin{tabular}{|c|cc|}
\hline
\multirow{2}{*}{Carrington Rotation} & \multicolumn{2}{c|}{Minimum Average Distance}                        \\ \cline{2-3} 
                                     & \multicolumn{1}{c|}{ADAPT-GONG} & GONG    \\ \hline
2106                                 & \multicolumn{1}{c|}{0.058}      & 0.108  \\ \hline
2123                                 & \multicolumn{1}{c|}{0.095}     & 0.113   \\ \hline
2137                                 & \multicolumn{1}{c|}{0.109}     & 0.141  \\ \hline
2154                                 & \multicolumn{1}{c|}{0.117}       & 0.092   \\ \hline
2167                                 & \multicolumn{1}{c|}{0.083}     & 0.121  \\ \hline
2174                                 & \multicolumn{1}{c|}{0.184}       & 0.193   \\ \hline
2198                                 & \multicolumn{1}{c|}{0.147}    & 0.173 \\ \hline
2209                                 & \multicolumn{1}{c|}{0.115}    & 0.147 \\ \hline
2222                                 & \multicolumn{1}{c|}{0.096}      & 0.125  \\ \hline
\end{tabular}
\caption{The minimum average distances for ADAPT-GONG and GONG magnetograms.}
\label{tab:dist}
\end{table}

\section{Summary and Discussions}

We use the GONG magnetograms to simulate the background solar wind with AWSoM in the
last solar cycle. 
We simulate the same Carrington rotations and 
follow the same methodology as described in Paper I. 
We only change the values of the Poynting flux parameter and use the
default values for all other parameters. We also apply the same
criteria in selecting the optimal Poynting flux parameter: 
the minimum value of the average of the curve distances between the simulated and observed density and the solar wind bulk velocity at 1\,AU.
This provides a direct comparison between simulations using ADAPT-GONG and GONG magnetograms, respectively.

There are several similarities between the results obtained from ADAPT-GONG magnetograms and GONG magnetograms. First
of all, the optimal Poynting flux parameter is also linearly correlated with 
the open field area and anti-correlated with the average unsigned $B_r$ in the open regions,
even though the Spearman's correlation coefficients are slightly less for GONG magnetogram:
it drops from 0.96 to 0.75 for the open field area and drops from $-0.91$ to $-0.79$ for the 
average unsigned $B_r$ in the open regions. Another similarity is that the average energy 
deposition rates from the two types magnetograms are almost the same.
Performing a similar study with another first-principle
solar wind model, such as the Magnetohydrodynamic Algorithm outside a Sphere 
(MAS, \cite{Mikic_1999}) and the MHD model based on Reynolds-averaged solar wind equations \citep{Usmanov_2018}, could help the community understand further if this conclusion holds 
for the Alfv\'en wave dissipation theory implemented into AWSoM, or this is generally
true for other similar first-principle based solar wind models. Another direction to validate 
this finding is to derive the energy deposition rate in 
open field regions from observations during
different phases of the solar cycle. There are some studies 
along this direction. For example, \cite{de_Pontieu_2007} looked at 
the Alfv\'en wave energy at a specific period, while \cite{Morton_2019} 
studied Alfv\'enic waves in a solar cycle without
separating open and closed field regions.

We notice that simulated solar wind driven by the ADAPT-GONG magnetograms 
generally agrees better with OMNI data at 1\,AU than the results driven by GONG magnetograms.
For example, Figures~\ref{fig:diff_pf_2106} and \ref{fig:diff_pf_2137} show that the simulated solar wind speed (from GONG magnetograms) is much
larger than the observed solar wind speed, as compared to Figure~1 in Paper I.
This conclusion is confirmed by Table\,\ref{tab:dist}:
the minimum distances
from the ADAPT-GONG maps are smaller than the GONG maps, 
with only one exception
for CR2154 in 2014, which is near solar maximum. 
This exception could be due to the more complicated
flux transport process in a very active Sun than a relatively quiet Sun,
and the ADAPT algorithm may be inaccurate
in that scenario.

Our study is an important step in using a first-principle solar wind model for
real-time solar wind prediction. Paper I and our results derived empirical
formulas for the optimal Poynting flux parameter of AWSoM, and we show 
AWSoM can produce reasonable solar wind predictions at
1\,AU, if the Poynting flux parameter is correctly set. 
Both Paper I and the current study cover only one Carrington
rotation per year between 2011-2019. 
It is to be checked if the empirical formulas will
hold if more rotations are included in the last solar cycle and/or other solar cycles are added.
These questions will be addressed in future work.

\begin{acknowledgments}

This work was primarily supported by the NSF PRE-EVENTS grant No. 1663800, 
the NSF SWQU grant No. PHY-2027555, the NSF Solar Terrestrial grant No. 2323303, 
and the NASA grants Nos. 80NSSC22K0269, 80NSSC23K0450 and 80NSSC22K0892.

We acknowledge the high-performance computing support from Cheyenne
(doi:10.5065/D6RX99HX) provided by NCAR's Computational and Information Systems Laboratory,
sponsored by the NSF, and the computation time on Frontera (doi:10.1145/3311790.3396656)
sponsored by NSF and the NASA supercomputing system Pleiades. 

This work utilizes data produced collaboratively between AFRL/ADAPT and NSO/NISP.

\end{acknowledgments}

\bibliography{reference}{}

\end{document}